\documentclass[
  journal=pasa,
  manuscript=research-paper, 
  year=2023,
  volume=37,
]{cup-journal}

\usepackage{microtype,siunitx,booktabs}
\usepackage{graphicx}
\usepackage{hyphenat}
\usepackage{amssymb}

\title{Evolved galaxies in high-density environments across $2.0\leq z<4.2$ using the ZFOURGE survey}

\author{Georgia R. Hartzenberg}
\affiliation{School of Chemistry and Physics, Queensland University of Technology, Brisbane, QLD 4000, Australia}
\email[Georgia R. Hartzenberg]{grh8893@gmail.com}

\author{Michael J. Cowley}
\affiliation{School of Chemistry and Physics, Queensland University of Technology, Brisbane, QLD 4000, Australia}
\alsoaffiliation{Centre for Astrophysics, University of Southern Queensland, Toowoomba, QLD 4350, Australia}

\author{Andrew M. Hopkins}
\affiliation{Australian Astronomical Optics, Macquarie University, North Ryde, NSW 2113, Australia}

\author{Rebecca J. Allen}
\affiliation{Centre for Astrophysics and Supercomputing, Swinburne University of Technology, Hawthorn, VIC 3122, Australia}

\received {dd Mmm YYYY}
\revised  {dd Mmm YYYY}
\accepted {dd Mmm YYYY}
\published{Day Month Year}

\keywords{} 

\begin{document}

\begin{abstract}
To explore the role environment plays in influencing galaxy evolution at high redshifts, we study $2.0\leq z<4.2$ environments using the FourStar Galaxy Evolution (ZFOURGE) survey. Using galaxies from the COSMOS legacy field with ${\rm log(M_{*}/M_{\odot})}\geq9.5$, we use a seventh nearest neighbour density estimator to quantify galaxy environment, dividing this into bins of low, intermediate and high density. We discover new high density environment candidates across $2.0\leq z<2.4$ and $3.1\leq z<4.2$. We analyse the quiescent fraction, stellar mass and specific star formation rate (sSFR) of our galaxies to understand how these vary with redshift and environment. Our results reveal that, across $2.0\leq z<2.4$, the high density environments are the most significant regions, which consist of elevated quiescent fractions, ${\rm log(M_{*}/M_{\odot})}\geq10.2$ massive galaxies and suppressed star formation activity. At $3.1\leq z<4.2$, we find that high density regions consist of elevated stellar masses but require more complete samples of quiescent and sSFR data to study the effects of environment in more detail at these higher redshifts. Overall, our results suggest that well-evolved, passive galaxies are already in place in high density environments at $z\sim2.4$, and that the Butcher-Oemler effect and SFR-density relation may not reverse towards higher redshifts as previously thought.
\end{abstract}

\section{INTRODUCTION}
\label{sec:int}
Galaxies are some of the most complex bodies in the universe. With different structural features such as bulges, arms, disks, bars, tidal tails and warps, the apparent `zoo' of optical morphologies in the local universe has long motivated questions about how galaxies have evolved over cosmic time. The processes of galaxy formation and evolution appear to depend on the proximity of neighbouring galaxies, more commonly referred to as `environment'. Galaxy environments, such as groups and clusters, are some of the densest regions in the universe and represent an example of external processes responsible for accelerating galaxy evolution. Galaxies in overdense environments are generally thought to have their evolutions influenced by (1) large-scale galaxy-galaxy interactions like merger activity \citep[e.g.][]{lin_redshift_2008,lin_where_2010,jian_pan-starrs1_2017} and harassment processes \citep[e.g.][]{moore_galaxy_1996,moore_morphological_1998,dutta_turbulence_2010}, or (2) hydrodynamical interactions like ram pressure stripping \citep[e.g.][]{gunn_infall_1972,abadi_ram_1999,fujita_rampressure_2001,hester_ram_2006}, viscous stripping \citep[e.g.][]{nulsen_transport_1982} and other gas depleting processes \citep[e.g.][]{bekki_passive_2002,kawata_strangulation_2008,peng_strangulation_2015,maier_clash-vlt_2016}. Early galaxy environment studies \citep[e.g.][]{oemler_systematic_1974,dressler_galaxy_1980,postman_morphology-density_1984} found that elliptical or bulge-like morphologies dominate regions like dense cluster cores; a correlation now known as the \textit{morphology-density relation}, while other work has argued that environment affects other galaxy properties like age \citep[e.g.][]{deng_environmental_2020} and star formation activity \citep[e.g.][]{hashimoto_influence_1998,cooper_deep2_2007,vulcani_comparing_2010,rasmussen_suppression_2012}.

While galaxy environments in the low redshift universe have been studied extensively, the role that environment plays in high redshift galaxy evolution remains poorly understood. For example, there is strong evidence that quiescent or passive galaxies are preferentially found in dense environments towards lower redshifts \citep[e.g.][]{wijesinghe_galaxy_2012,lin_pan-starrs1_2014,damjanov_environment_2015,davies_galaxy_2016,allen_differences_2016,jian_pan-starrs1_2017}, but at $z\sim1-1.5$ this trend appears to reverse \citep[e.g.][]{lin_splash_2016}. This effect is known as the Butcher-Oemler effect \citep[][]{butcher_evolution_1978}, which states that the fraction of blue, star-forming galaxies in overdense environments increases with redshift. However, some high redshift measurements \citep[e.g.][]{allen_differences_2016,kawinwanichakij_effect_2017} disagree, where comparisons between low and high density environments reveals the latter to host a larger fraction of red, passive galaxies.

Furthermore, there is evidence to suggest that the relationship between passive galaxy fractions and environment is closely correlated with the relation between star formation rate (SFR) and density \citep[e.g.][]{bolzonella_tracking_2010,peng_mass_2010,deng_environmental_2011,lu_cfht_2012}, whereby an increase in passive galaxy fractions drives the suppression of star formation in dense environments in the local universe. Analogous to the work of \citet{butcher_evolution_1978}, some studies \citep[e.g.][]{elbaz_reversal_2007,cooper_deep2_2007} have argued that $z\sim1$ represents a `transition point', where regions of high density instead have enhanced star formation activity. In support of this picture, \citet{patel_star-formation-rate-density_2011} find low star formation activity and high quiescent fractions in groups and a cluster at redshifts $0.6<z<0.9$ while at higher redshifts \citep[e.g. $z\geq1.5$; ][]{tran_reversal_2010,hayashi_high_2010,hayashi_properties_2011,strazzullo_galaxy_2013,santos_reversal_2015,wang_discovery_2016}, overdense environments tend to have enhanced star formation activity. This picture is not clear-cut though, with some work finding that $z>1$ high density environments consist of suppressed star formation activity when compared to lower density counterparts \citep[e.g.][]{lidman_hawk-i_2008,grutzbauch_relationship_2011,old_gogreen_2020}. Other studies \citep[e.g.][]{darvish_effects_2016} have even found that star formation activity does not significantly change between environments out to $z\sim3$, and that environmental quenching is strongest towards lower redshifts \citep[i.e. $z\leq1$; also suggested by][]{kawinwanichakij_effect_2017}. 

Finally, the $z\sim2-3$ regime is considered a key epoch, sometimes called `cosmic noon' \citep[][]{schreiber_star-forming_2020}, where universal star formation peaked, galaxies formed approximately half of their stellar mass, the dusty star-forming population peaked
\citep[e.g.][]{simpson_alma_2014,dudzeviciute_alma_2020} and AGN activity was the most prominent \citep[][]{cowley_decoupled_2018}. In addition, there is strong evidence that merger interactions were common in the early universe \citep[e.g.][]{lin_redshift_2008,tacconi_submillimeter_2008,allen_differential_2015,li_merging_2022}, and that the apparent growth of stellar mass \citep[e.g.][]{tomczak_galaxy_2014} may be driven by these merger processes \citep[e.g.][]{conselice_direct_2003,naab_minor_2009,van_dokkum_growth_2010}. The `cosmic noon' consequently encompasses some of the most significant stages of galaxy evolution.

To determine how the relationship between quiescent galaxy fraction, star formation activity and environmental density evolves over cosmological redshift, we take advantage of the FourStar Galaxy Evolution (ZFOURGE) survey and study the properties of environment spanning redshifts of $2.0\leq z<4.2$. This paper is structured as follows. In \S~\ref{sec:samples}, we introduce the ZFOURGE catalogues and the parameters we employ for this study. In \S~\ref{sec:methods}, we outline our methods for creating our subsamples, searching for overdense environments and defining low, intermediate and high environmental density. In \S~\ref{sec:res}, we present our comparative analysis of the environments and discuss the implications of the results in \S~\ref{sec:discussion}. We then summarise our main findings and provide concluding remarks in \S~\ref{sec:summary}. The work in this study assumes a $\Lambda$CDM cosmology of $H_{0}=70\,$km\,s$^{-1}$\,Mpc$^{-1}$, $\Omega_{m}=0.3$ and $\Omega_{\Lambda}=0.7$.

\section{THE SAMPLES}
\label{sec:samples}

\subsection{ZFOURGE catalogues}
\label{subsec:zfourge}
The parent sample of this work consists of galaxies from the $2017$ release of the ZFOURGE\footnote{https://zfourge.tamu.edu} survey \citep{straatman_fourstar_2016}. ZFOURGE consists of approximately $60,000$ galaxies at $z>0.1$, and was taken using the near-IR FourStar imager \citep{persson_fourstar_2013}, mounted on the $6.5$-metre Magellan Baade Telescope at the Las Campanas Observatory, Chile. The ZFOURGE survey covered three $11' \times 11'$ HST legacy fields, COSMOS \citep{scoville_cosmic_2007}, CDFS \citep{giacconi_chandra_2002} and UDS \citep{lawrence_ukirt_2007}, and employed deep near-IR imaging in the $J_{1}$, $J_{2}$, $J_{3}$, $H_{l}$ and $H_{s}$ medium-band filters and $K_{s}$ broad-band filter. The imaging of ZFOURGE, measured over $~1.0-1.8\mu$m, consists of a circular aperture of diameter $D = 0.6"$ and reaches $5\sigma$ point-source limited depths of $~26$ AB mag in the $J$ medium-bands and $~25$ AB mag in the $H$ and $K_{s}$-bands \citep{spitler_first_2012}. For sources across $1<z<4$, these filters result in well-constrained photometric redshifts with $\Delta z/(1+z) \simeq 1-2\%$ \citep{straatman_fourstar_2016} and better detection of quiescent and dusty star-forming sources. ZFOURGE is also supplemented with existing public data from the HST/WFC3 F160W and F125W imaging of the CANDELS \citep{grogin_candels_2011, koekemoer_candels_2011} survey and also contains data from the \textit{Spitzer}/Infrared Array Camera (IRAC) and \textit{Herschel}/Photodetector Array Camera and Spectrometer (PACS). For a full description of the data and methods of ZFOURGE, see \citet{straatman_fourstar_2016}.

\subsection{Photometric redshifts, UVJ rest-frame colour diagram, stellar mass, specific star formation rates and AGN candidates}
\label{subsec:zfourge_properties}
In this study, we use the photometric redshifts, rest-frame colours, stellar masses and star formation rates of the COSMOS legacy field of ZFOURGE. Photometric redshifts are determined using \texttt{EAZY} \citep{brammer_eazy_2008}, which fits linear combinations of nine spectral templates to the $0.3-8\mu$m observed photometry of the galaxies. Five of the templates are from the library of \texttt{P\'{E}GASE} stellar population synthesis models \citep{fioc_pegase_1997}. The remaining four templates are that of young and dusty galaxies, old reddened galaxies, old dusty galaxies and galaxies with strong emission lines. The ZFOURGE collaboration also chose to include a template error function to account for wavelength-dependent uncertainties. Offsets in zeropoints, which may affect flux and therefore the photometric redshifts, are corrected for by iteratively fitting the \texttt{EAZY} templates to the SEDs of the ZFOURGE galaxies. During the fitting procedure, the ZFOURGE collaboration calculated the systematic offsets between the templates and data, and modified the templates accordingly. This allows the templates to highlight fainter features of galaxies including dust-absorption at $2175$\AA. Zeropoint corrections are then calculated in the observed frame. This fitting process was repeated until the zeropoint corrections in the bands were $<1\%$, which was typically after three or four iterations. Lastly, the ZFOURGE photometric redshifts were compared to spectroscopic redshifts in the same fields provided by the 3D-HST survey \citep{skelton_3d-hst_2014}. The ZFOURGE collaboration used the Normalised Median Absolute Deviation (i.e. $\sigma_{z} = 1.48 \times median$ $absolute$ $value$ $[\mid z_{phot}-z_{spec}\mid/(1+z_{spec})]$) to quantify the errors and achieved excellent results (COSMOS $\sigma_{z}=0.009$). For more information on the photometric redshifts, refer to Section 5 of \citet{straatman_fourstar_2016}.

We use the ${\rm UVJ}$ rest-frame colour diagram to divide the galaxy populations into quiescent, star-forming and dusty star-forming. Sources that reside in the upper-left region of the ${\rm UVJ}$ diagram are defined as being quiescent, given by the vertices $({\rm V}-{\rm J},{\rm U}-{\rm V})=(-\infty,1.3),(0.85,1.3),(1.6,1.95),(1.6,+\infty)$, while the vertical boundary of $({\rm V}-{\rm J})=1.2$ divides the star-forming and dusty star-forming populations \citep{spitler_exploring_2014}. The diagram has been shown to be an efficient way of separating the quiescent and star-forming populations \citep[e.g.][]{wuyts_what_2007,williams_detection_2009,wild_new_2014}.

The stellar masses of ZFOURGE are determined by fitting the stellar population synthesis models of \citet{bruzual_stellar_2003} using \texttt{FAST} \citep{kriek_ultra-deep_2009}. The models assume solar metallicity, the $A_{v} = 0-4$ dust extinction law of \citet{calzetti_dust_2000}, IMF of \citet{chabrier_galactic_2003} and an exponentially decreasing star formation history.

The ZFOURGE SFRs consider the rest-frame UV emission from massive stars and the IR emission re-radiated from dust-obscured stars. SFRs are determined via methods detailed in \citet{bell_toward_2005}, scaled to the IMF of \citet{chabrier_galactic_2003}, and given by, 
\begin{equation}
    \Psi_{IR+UV}[{\rm M_\odot}{\rm yr^{-1}}]=1.09\times 10^{-10}(3.3L_{UV}+L_{IR}),
\end{equation}
where $L_{UV}$ is the \texttt{EAZY}-derived rest-frame luminosity, integrated over $1216-3000$\AA, and $L_{IR}$ is the bolometric IR luminosity, integrated over $8-1000\mu$m and calculated via a luminosity-independent conversion \citep{wuyts_fireworks_2008,wuyts_star_2011} using \textit{Herschel}/PACS and \textit{Spitzer}/MIPS out to $160\mu$m fluxes where available. The inclusion of IR emission results in more robust SFRs \citep{tomczak_sfr_2016}. To eliminate the bias of stellar mass, we use the specific star formation rate (sSFR) parameter, which is a relative measure of the level of star formation activity within a galaxy. The sSFRs of ZFOURGE (in units of ${\rm Gyr^{-1}}$) are calculated by dividing the SFRs by stellar mass.

Finally, we highlight that the ZFOURGE catalogues also include radio, IR and X-ray AGN data \citep{cowley_zfourge_2016}. For our study, we use these three bands to generate a simple data set that indicates the AGN status of the galaxies, where positive AGN detection $= 1$, no AGN detection $= 0$. This new data is used to reduce contamination in the sSFRs (\S~\ref{subsec:ssfr}), and to highlight the on-sky positions of AGN-dominated galaxies in the results of \S~\ref{subsec:7NN_maps}. According to \citet{cowley_zfourge_2016}, the ZFOURGE AGN candidates have minimal impact on the derived photometric redshifts, SFRs and UVJ colours. However, given that AGN galaxies are notorious for contaminating light from star formation activity \citep[e.g.][]{juneau_widespread_2013}, we decided to take a cautious approach and omit the AGN candidates from our sSFR samples. The majority of the ZFOURGE AGN ($\sim 85\%$) have been selected via IR techniques and are classified as Type 2. This is because Type 2 sources are more readily detected via IR selection due to the presence of obscuring material. $\sim 90\%$ of the AGN sources in this study exhibit characteristics which are consistent with Type 2 AGN. For more information on the ZFOURGE AGN classification methods, refer to \citet{cowley_zfourge_2016}. 
\\
\\

\section{GALAXY ENVIRONMENT SEARCH}
\label{sec:methods}

\subsection{Mass-limited sample, redshift slices and 7NN algorithm}
\label{subsec:gal_env_search}
For our galaxy sample, we isolate COSMOS sources with \texttt{use=1} flags that lie in $2.0\leq z< 4.2$. The \texttt{use} flag provides a standard selection of galaxies that have been surveyed by all medium-band filters of ZFOURGE, and consist of good photometry \citep{straatman_fourstar_2016}. The distribution of stellar masses within the ZFOURGE survey is not consistent, with lower mass galaxies (${\rm log(M_{*}/M_{\odot})}<8$) being more prominent towards $z<1$ while higher redshifts consist of a bias towards larger stellar masses. To reduce this bias and obtain a more complete sample, we impose a stellar mass limit of ${\rm log(M_{*}/M_{\odot})}\geq9.5$ on our $2.0\leq z< 4.2$ COSMOS sources. This mass cut is consistent with the $80\%$ stellar mass completeness limits of \citet{tomczak_galaxy_2014} and \citet{papovich_zfourgecandels_2015}.

We define eight redshift slices: $2.0\leq z<2.2$ $(z_{1})$, $2.2\leq z<2.4$ $(z_{2})$, $2.4\leq z<2.6$ $(z_{3})$, $2.6\leq z<2.8$ $(z_{4})$, $2.8\leq z<3.1$ $(z_{5})$, $3.1\leq z<3.4$ $(z_{6})$, $3.4\leq z<3.8$ $(z_{7})$ and $3.8\leq z<4.2$ $(z_{8})$. We tested the widths of our redshift slices by comparing comoving space densities in the form of $N/V_{C}$, where $N$ and $V_{C}$ is the galaxy count and comoving volume of a given redshift slice, respectively. We find that the comoving space densities decrease as redshift increases, which in turn reflects the distribution of the COSMOS galaxies in ZFOURGE. We also tested larger redshift slice widths (i.e. $\delta z\leq0.6$) and found that the qualitative result remains unchanged. Lastly, we examined the photometric redshift uncertainties of ZFOURGE and found that, across bins of $2.0<z<3.0$ and $3.0<z<4.0$, the average redshift errors are $\sim0.146$ and $\sim0.172$, respectively. Our redshift slice selection, therefore, accommodates the uncertainties in the photometric redshifts.

We searched for high density environments over the eight redshift slices by generating projected surface density maps using the seventh nearest neighbour (7NN) metric \citep{spitler_first_2012}. At each `pixel' on the ZFOURGE COSMOS sky, we calculated the projected surface density by isolating the seventh closest galaxy and its distance, given by,

\begin{equation} \label{eq:2}
    n_{7}=\frac{N}{\pi r_{N}^2},
\end{equation}
where $r_{N}$ is the on-sky distance to the $\rm n^{th}$ closest galaxy and $N=7$. The pixels range from the minimum to maximum RA and Dec coordinates of the galaxies, and have an adopted spacing of $0.000357$ deg $(\sim 1.29")$. Finally, to quantify the statistical significance of the high density environments, the 7NN densities of each redshift slice are averaged using the means and standard deviations of the adjacent redshift slices (e.g. for $z_{1}$, adjacent slices are $1.8\leq z<2.0$ and $2.2\leq z<2.4$.). The 7NN density estimates for the adjacent redshift slices are evaluated only at the locations of the galaxies. To show contrast between colours, the 7NN densities are plotted via a logarithmic scale, over a 0.1-20 range. We selected redshift slices $z_{1}$, $z_{2}$, $z_{6}$ and $z_{8}$ for our study, elaborate on this choice in \S~\ref{subsec:7NN_maps} and show the 7NN projected surface density maps of these slices in Figure \ref{7NN_maps}. 

These algorithms usually require testing to ensure that true overdensities are being observed and not imperfect photometric redshift measurements. \citet{spitler_first_2012} developed this 7NN metric and tested its reliability in two ways. First, they performed a bootstrap resampling of the photometric redshifts, shuffled the results of each iteration and produced new 7NN density maps. From this, it was found that only $3/1000$ resampled maps had one high density environment. For the second check, \citet{spitler_first_2012} generated 121 mock density maps using the light cone catalogues of the online laboratory Mock Galaxy Factory \citep{bernyk_theoretical_2016}, and found that the mock maps had a scatter which was consistent with that of the real density map. The above results indicate that the 7NN overdensities are not random associations and are therefore robust and reliable. \citet{allen_differential_2015}, who studied the $z_{1}$ H2 and H3 overdensities in detail, also carried out tests for the 7NN algorithm and achieved similar results. Different values of $N$ (i.e. $N=5-9$) in Equation \ref{eq:2} were tested by \citet{spitler_first_2012}, who confirmed that the results do not significantly change for the surface density maps.

\subsection{Environmental density definitions}
\label{subsec:env_def}
\begin{figure*}[ht!]
\centering
\includegraphics[width=0.975\textwidth]{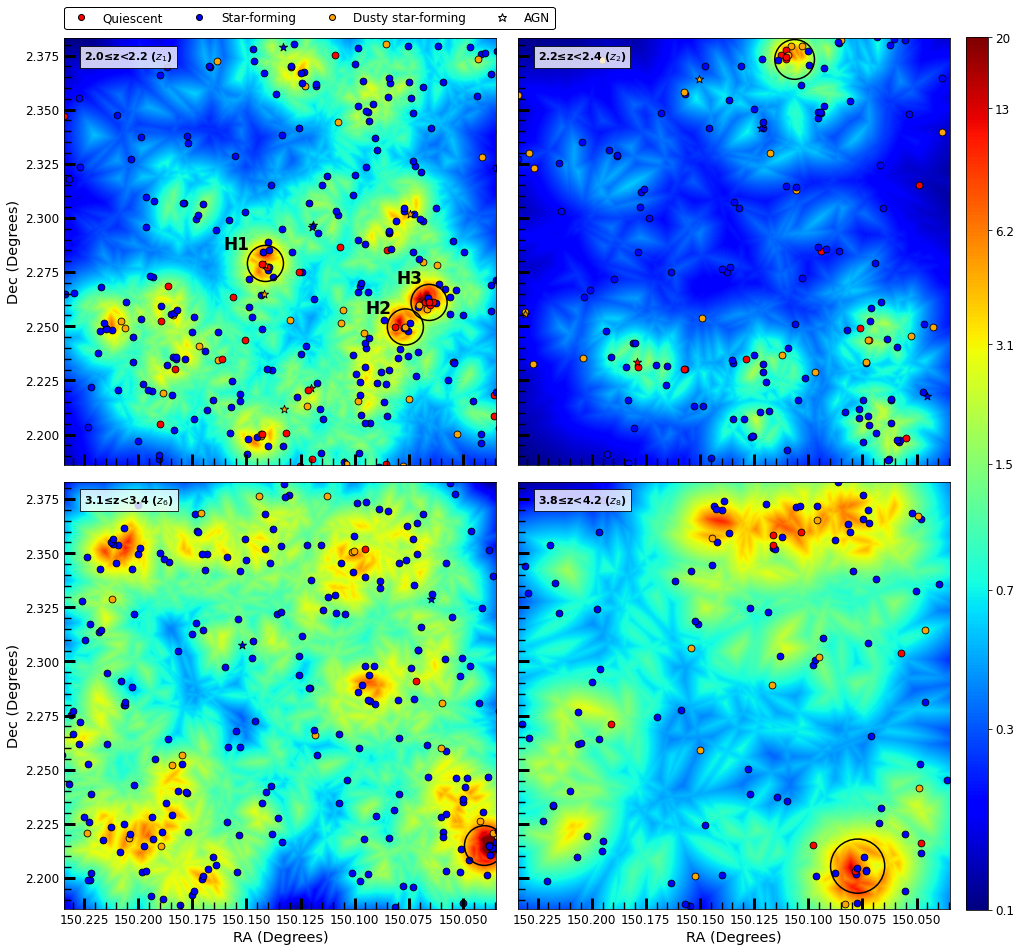}
\caption{Projected surface density maps of COSMOS redshift slices $2.0\leq z<2.2$ ($z_{1}$; upper-left), $2.2\leq z<2.4$ ($z_{2}$; upper-right), $3.1\leq z<3.4$ ($z_{6}$; lower-left) and $3.8\leq z<4.2$ ($z_{8}$; lower-right). The general location of the high density environments are shown by the apertures (black circles). The $z_{1}$, $z_{2}$, $z_{6}$ and $z_{8}$ apertures are $\sim$ $30"$, $33"$, $33"$ and $45"$, respectively. The colour bar shows the statistical significance of the 7NN densities above the mean density averaged over adjacent redshift slices. H2 and H3 in $z_{1}$ represent the original overdensities identified by \citet{spitler_first_2012} and spectroscopically confirmed by \citet{yuan_keckmosfire_2014}. Quiescent, star-forming and dusty star-forming sources are shown as red, blue and orange markers, respectively, while those that have an AGN according to the catalogues of \citet{cowley_zfourge_2016} are shown as star markers.}
\label{7NN_maps}

\end{figure*}

\begin{figure*}[ht!]
    \centering
    \includegraphics[width=0.94\textwidth]{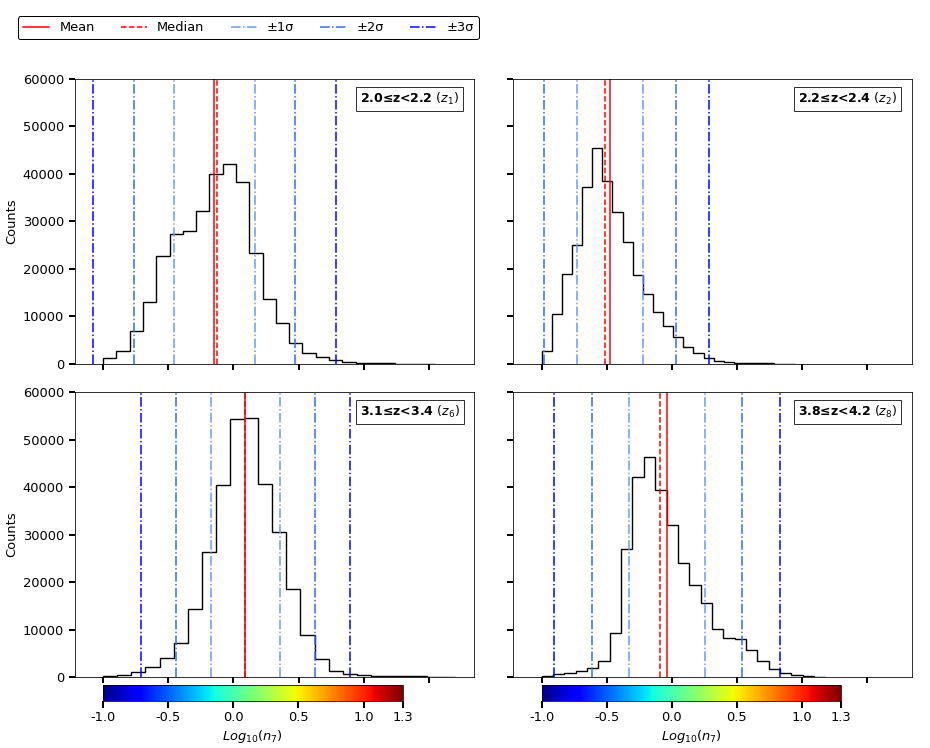}
    \caption{Distributions of the logarithmic-scaled 7NN surface density pixels of $2.0\leq z<2.2$ ($z_{1}$; upper-left), $2.2\leq z<2.4$ ($z_{2}$; upper-right), $3.1\leq z<3.4$ ($z_{6}$; lower-left) and $3.8\leq z<4.2$ ($z_{8}$; lower-right). The colour bar corresponds to that of Figure \ref{7NN_maps}. Means and medians of the densities are shown as solid and dashed red lines, respectively, while $\pm\sigma$, $\pm2\sigma$ and $\pm3\sigma$ are given by the darkening dot-dash blue lines.}
    \label{env_def}
\end{figure*}

Figure \ref{env_def} shows the distributions of the 7NN projected surface density pixels of $z_{1}$, $z_{2}$, $z_{6}$ and $z_{8}$ and includes the colour bar of Figure \ref{7NN_maps} for comparison. The means, medians, $\pm\sigma$, $\pm2\sigma$ and $\pm3\sigma$ parameters are shown as the solid red, dashed red, and darkening blue dot-dashed lines, respectively. We chose a fixed threshold in $\log(n_{7})$ below, between and above $-0.5$ and $+0.5$ to define low, intermediate and high density, respectively. We therefore selected the $\sigma$-valued boundaries of Table \ref{env_table} to most closely approximate the $\log(n_{7})$ environmental density threshold. We tested the $25^{th}$ and $75^{th}$ quartiles, but these parameters were not symmetrical around the non-Gaussian distributions and overestimated the low and high density environments due to the very uneven cuts. Different definitions using the $\sigma$ parameters were also tested, but we concluded that those of Table \ref{env_table} divided the sources into low, intermediate and high density environments most consistently. Previous ZFOURGE environment papers that employed the same 7NN algorithm \citep[e.g.][]{spitler_first_2012, allen_differential_2015, allen_differences_2016} also referred to their overdensities in terms of $\sigma$. We provide the environment sample sizes of the redshift slices in Table \ref{sample_table}.

\begin{table*}[h!]
\centering
\resizebox{0.80\textwidth}{!}{%
\begin{tabular}{llll}
\hline
\multicolumn{4}{c}{\cellcolor[HTML]{FFFFFF}\textbf{Environmental Density}}   \\ \hline
\textbf{Redshift Slice}                     & \textbf{Low}    & \textbf{Intermediate} & \textbf{High}   \\ \hline
\textbf{$2.0\leq z<2.2$ $(z_{1})$} 
& ${\rm log_{10}}(n_{7})<-\sigma$ 
& $-\sigma\leq {\rm log_{10}}(n_{7})<2\sigma$ 
& ${\rm log_{10}}(n_{7})\geq2\sigma$  \\ \hline 

\textbf{$2.2\leq z<2.4$ $(z_{2})$} 
& ${\rm log_{10}}(n_{7})\leq\sigma$ 
& $\sigma<{\rm log_{10}}(n_{7})\leq3\sigma$ 
& ${\rm log_{10}}(n_{7})>3\sigma$  \\ \hline 

\textbf{$3.1\leq z<3.4$ $(z_{6})$} 
& ${\rm log_{10}}(n_{7})<-2\sigma$ 
& $-2\sigma\leq {\rm log_{10}}(n_{7})<2\sigma$ 
& ${\rm log_{10}}(n_{7})\geq2\sigma$  \\ \hline 

\textbf{$3.8\leq z<4.2$ $(z_{8})$} 
& ${\rm log_{10}}(n_{7})<-\sigma$ 
& $-\sigma\leq {\rm log_{10}}(n_{7})<2\sigma$ 
& ${\rm log_{10}}(n_{7})\geq2\sigma$ \\ \hline 
\end{tabular}%
}
\caption{Environmental density definitions of the four redshift slices, derived from the 7NN density distributions of Figure \ref{env_def}. The low density $\sigma$-boundaries of $z_{1}$, $z_{2}$, $z_{6}$ and $z_{8}$ are $\sim$ $-0.5$, $-0.2$, $-0.4$ and $-0.3$, respectively. The high density $\sigma$-boundaries of $z_{1}$, $z_{2}$, $z_{6}$ and $z_{8}$ are $\sim$ $0.5$, $0.3$, $0.6$ and $0.5$, respectively.}

\label{env_table}
\end{table*}

\section{RESULTS}
\label{sec:res}

\subsection{Projected surface density maps}
\label{subsec:7NN_maps}
We discover new high density environment candidates in the COSMOS legacy field within redshift slices $z_{1}$, $z_{2}$, $z_{6}$ and $z_{8}$ as shown by the apertures (black circles) in the 7NN projected surface density maps of Figure \ref{7NN_maps}. $z_{1}$ apertures are $\sim 30"$, $z_{2}$ and $z_{6}$ apertures are $\sim 33"$ while the $z_{8}$ aperture is $\sim 45"$. The colour bar represents the statistical significance (in $\sigma$) of the 7NN densities above the mean density, which is averaged over the adjacent redshift slices. In Figure \ref{7NN_maps}, H2 and H3 represent the overdensities found by \citet{spitler_first_2012} and spectroscopically confirmed by \citet{yuan_keckmosfire_2014}, while H1 appears to be a new high density candidate for this slice. Redshift slices $z_{2}$, $z_{6}$ and $z_{8}$ have one newly detected high density environment each. We did not identify any concentrated ${\rm log_{10}}(n_{7})\gtrsim+0.5$ high density environments across $z_{3}$, $z_{4}$, $z_{5}$ and $z_{7}$, and so omit these redshift slices from this study. 

\subsection{Quiescent fractions}
\label{subsec:gal_frac}
\begin{figure*}[h]
    \centering
    \includegraphics[width=0.8\textwidth]{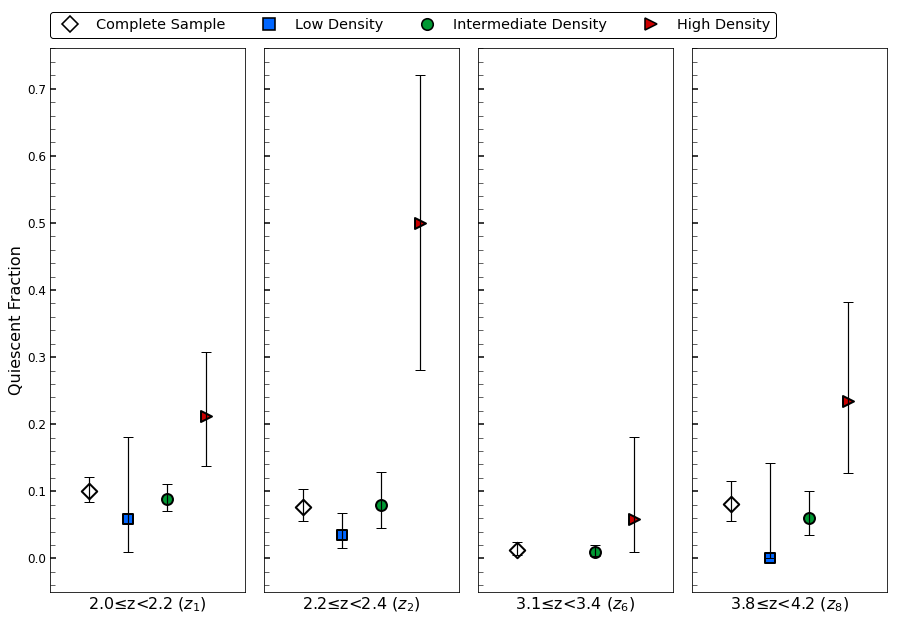}
    \caption{Quiescent galaxy fractions of the environments and complete samples of $2.0\leq z<2.2$ $(z_{1})$, $2.2\leq z<2.4$ $(z_{2})$, $3.1\leq z<3.4$ $(z_{6})$ and $3.8\leq z<4.2$ $(z_{8})$. Low, intermediate and high density environments are given by the blue square, green circle and red triangle markers, respectively, while the complete samples are represented by the white diamonds. Errors shown indicate $1\sigma$ Clopper-Pearson binomial confidence intervals. The quiescent fraction of the environments is given by $N_{q(env)}/N_{total(env)}$, while the quiescent fraction of the complete samples is in the form of $N_{q}/N_{total}$. The low density environment of $z_{6}$ is omitted due to low numbers.}
    \label{q_frac}
\end{figure*}

We compare the quiescent galaxy fractions as a function of environment and redshift. As described in \S~\ref{subsec:zfourge_properties}, the quiescent classification is based on the ${\rm UVJ}$ rest-frame colour diagram. For the complete samples, the quiescent fraction for a given redshift slice is in the form of $f_{q}=N_{q}/N_{total}$. For the environment samples, we find the count of a given galaxy population and divide this by the count of galaxies in that environment (e.g. low density quiescent fraction is defined as $f_{q(L)}=N_{q(L)}/N_{total(L)}$). Figure \ref{q_frac} shows the quiescent fractions of the low (blue squares), intermediate (green circles) and high density (red triangles) environments. Quiescent fractions of the complete samples (white diamonds) are also included for comparison. Errors shown are the $1\sigma$ Clopper-Pearson binomial confidence intervals. Only one source was identified as being part of the $z_{6}$ low density environment, and is therefore excluded from this analysis. Across all redshift slices, Figure \ref{q_frac} reveals that the quiescent galaxy fraction tends to increase with environmental density and appears to be largest for $z_{2}$. We note that $z_{6}$ and $z_{8}$ have significantly fewer quiescent sources than $z_{1}$ and $z_{2}$, and so caution is warranted with the results of Figure \ref{q_frac} at $z\geq3.1$. There is also some overlap in the uncertainties between the low and high density samples of $z_{1}$. Despite this, Figure \ref{q_frac} shows evidence that quiescent galaxies are preferentially found in high density environments across all redshifts probed in this study.

\subsection{Stellar mass}
\label{subsec:mass}

\begin{figure*}[ht!]
    \centering
    \includegraphics[width=0.8\textwidth]{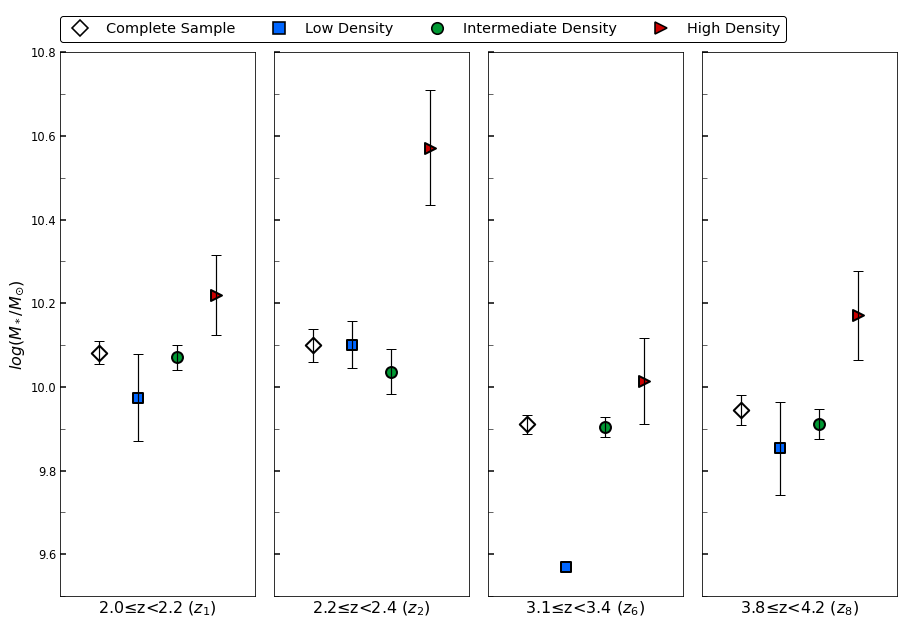}
    \caption{Average stellar mass, ${\rm log(M_{*}/M_\odot)}$, of the environments and complete samples of $2.0\leq z<2.2$ $(z_{1})$, $2.2\leq z<2.4$ $(z_{2})$, $3.1\leq z<3.4$ $(z_{6})$ and $3.8\leq z<4.2$ $(z_{8})$. Errors shown correspond to the $\rm {68\%}$ confidence intervals calculated from a bootstrap analysis. The stellar mass of the low density galaxy of $z_{6}$ is also included in the figure for completeness. Markers are as in Figure \ref{q_frac}.}
    \label{av_mass}
\end{figure*}

Figure \ref{av_mass} shows the average stellar mass as a function of environment and redshift. The stellar mass of the galaxy belonging to the $z_{6}$ low density environment is also included. Errors shown indicate the $\rm {68\%}$ confidence intervals evaluated via bootstrapping methods. For the bootstrapping analysis, we generated 9999 new samples from the original and calculated the average for each. All bootstrapped averages were then arranged in ascending order and the averages at the lower and upper percentiles were used to construct the lower and upper limits of the $68 \%$ confidence intervals. Across all redshift slices, Figure \ref{av_mass} reveals that the high density environments tend to have the most massive galaxies compared to the complete, low and intermediate density samples. This is strongest for redshift slices $z_{1}$ and $z_{2}$, where the average stellar masses for the high density environments were found to be ${\rm log(M_{*}/M_{\odot})}\sim10.220\pm0.096$ and $\sim10.571^{+0.139}_{-0.136}$, respectively. The average stellar masses of the $z_{6}$ and $z_{8}$ high density environments, on the other hand, were found to be ${\rm log(M_{*}/M_{\odot})}\sim 10.015^{+0.102}_{-0.104}$ and $\sim 10.172^{+0.106}_{-0.108}$, respectively. 

We found the low-high density stellar mass offsets for redshift slices $z_{1}$, $z_{2}$ and $z_{8}$ to be $\Delta\rm log(M_{*})\sim0.247^{+0.142}_{-0.141}$, $\sim0.470^{+0.149}_{-0.147}$ and $\sim0.318^{+0.154}_{-0.155}$ dex, respectively. For $z_{6}$, we calculated the high density-complete sample offset which evaluated to $\Delta\rm log(M_{*})\sim0.104^{+0.105}_{-0.107}$ dex. Redshift slices $z_{2}$ and $z_{8}$, therefore, have the most significant and greatest elevation of stellar mass.

\subsection{Star formation activity}
\label{subsec:ssfr}
\begin{figure}[h!]
    \centering
    \includegraphics[width=0.95\textwidth]{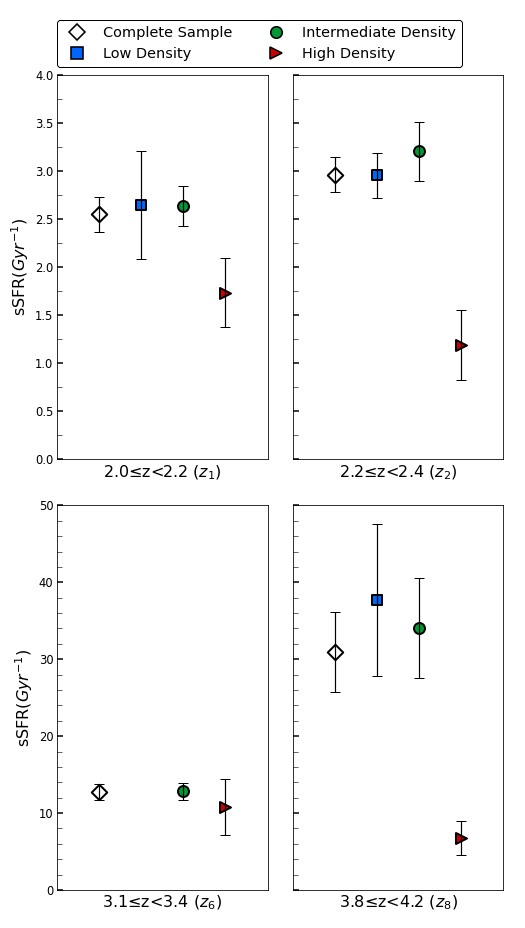}
    \caption{Average specific star formation rate (sSFR) in $\rm {Gyr^{-1}}$ of the environments and complete samples of $2.0\leq z<2.2$ $(z_{1})$, $2.2\leq z<2.4$ $(z_{2})$, $3.1\leq z<3.4$ $(z_{6})$ and $3.8\leq z<4.2$ $(z_{8})$. Redshift slices $z_{1}$ and $z_{2}$ share a horizontal axis, while $z_{6}$ and $z_{8}$ share another. Errors shown correspond to the $\rm {68\%}$ confidence intervals calculated from a bootstrap analysis. The $z_{6}$ low density galaxy is omitted from the figure due to unavailable star formation data. AGN candidates of \citet{cowley_zfourge_2016} are also excluded from the sSFR samples. Markers are as in Figure \ref{q_frac}.}
    \label{av_ssfr}
\end{figure}

Figure \ref{av_ssfr} shows the average sSFRs as a function of environment and redshift. Errors shown are the $\rm {68\%}$ confidence intervals evaluated via bootstrapping methods described in \S~\ref{subsec:mass}. We found that the sSFR ranges differed between the lower and higher redshift slices. Therefore, to better highlight the contrast in star formation activity, we pair $z_{1}$ and $z_{2}$ on one axis and $z_{6}$ and $z_{8}$ on another. As mentioned in \S~\ref{subsec:zfourge_properties}, we remove the \citet{cowley_zfourge_2016} AGN candidates from the sSFR samples to reduce contamination. The low density environment of $z_{6}$ lacks SFR data and so is excluded from this analysis. From Figure \ref{av_ssfr}, we find evidence that the high density environments tend to have the lowest sSFRs across all redshift slices. This is particularly strong for $z_{8}$.

We find that from low to high density, the average sSFRs decrease by $\sim34\%$, $\sim60\%$ and $\sim82\%$ for redshift slices $z_{1}$, $z_{2}$ and $z_{8}$, respectively. For $z_{6}$, we determined the percentage difference between the complete sample and high density environment, which is $\sim15\%$. Therefore, we find that the drop in star formation activity is greatest across redshift slices $z_{2}$ and $z_{8}$.

\begin{table}[h!]
\centering
\resizebox{\textwidth}{!}{%
\begin{tabular}{llll}
\hline
\multicolumn{4}{r}{\cellcolor[HTML]{FFFFFF}\textbf{Environmental Density}}   \\ \hline
\textbf{Redshift Slice}   & \textbf{Low} & \textbf{Intermediate} & \textbf{High} \\ \midrule
$2.0\leq z<2.2$ $(z_{1})$ & $17$           & $247$                   & $33$           \\ \midrule
$2.2\leq z<2.4$ $(z_{2})$ & $86$           & $63$                   & $8$             \\ \midrule
$3.1\leq z<3.4$ $(z_{6})$ & $1$            & $223$                   & $17$            \\ \midrule
$3.8\leq z<4.2$ $(z_{8})$ & $12$           & $82$                    & $17$           \\ \bottomrule
\end{tabular}%
}
\caption{Environment sample sizes of the four redshift slices, according to the definitions of Table \ref{env_table}.}

\label{sample_table}
\end{table}

\section{DISCUSSION}
\label{sec:discussion}
For the first time using ZFOURGE data, we have studied the influence of environment on galaxy evolution spanning redshifts of $2.0\leq z<4.2$. With previous work debating the redshift at which the Butcher-Oemler effect reverses (and if it does at all), we compared the quiescent galaxy fractions and average stellar masses and sSFRs in regions of low, intermediate and high density in the early universe. Predominately, it appears that high density environments contain evolved, passive galaxies as far back as $z\sim2.4$ and potentially to the much higher redshift of $z\sim 4$.

From Figure \ref{q_frac}, we find that the quiescent galaxy fraction tends to increase with environmental density across all redshift slices probed in this study. It should be noted, however, that the quiescent galaxy numbers naturally decrease as photometric redshift increases within the ZFOURGE catalogues, meaning that we may not have enough quiescent systems at $z\geq3.1$ to make a fair comparison between environments. Reinforcing our results, though, is that they are consistent with those of \citet{strazzullo_galaxy_2013}, for example, who also found that a dense cluster core at $z\sim2$ had an enhanced passive galaxy fraction. Work by \citet{spitler_first_2012}, who studied the photometrically identified H2 and H3 overdensities of $z_{1}$, also highlighted that galaxies in these environments tended to be redder than the field or low density population. The results of Figure \ref{q_frac} appear to suggest that the positive correlation between passive galaxy fractions and environmental density in the low redshift universe \citep[e.g. $z<2$;][]{dressler_galaxy_1980,lidman_hawk-i_2008,damjanov_environment_2015,allen_differences_2016,davies_galaxy_2016,jian_pan-starrs1_2017,kawinwanichakij_effect_2017} extends into the higher redshift universe, to at least $z\sim2.4$.

Figure \ref{av_ssfr} reveals that, across all redshift slices, the high density environments have the lowest sSFRs. From low to high density, we found that the $z_{1}$ star formation activity decreased by $\sim 34\%$, while in $z_{2}$ star formation activity dropped by $\sim 60\%$. Coupled with our quiescent fraction findings, the suppressed sSFRs provide evidence to suggest that the SFR-density relation and Butcher-Oemler effect do not appear to reverse at higher redshifts. Our findings are in direct contrast to those of \citet{wang_discovery_2016}, who found that a dense concentration of massive galaxies at $z=2.506$ was dominated by star-forming sources and high SFRs of $\sim3400$ ${\rm M_\odot yr^{-1}}$ as well as other high redshift studies \citep[e.g. $z\geq1.2$;][]{tran_reversal_2010, hayashi_properties_2011,alberts_evolution_2014,santos_reversal_2015} who find that their high density environments experienced accelerated levels of star formation activity.

The question now is why we observe well-evolved, passive sources in overdense environments across $2.0\leq z<2.4$. With the fates of galaxies generally controlled by their stellar masses \citep[e.g.][]{peng_mass_2010,rasmussen_suppression_2012,davies_galaxy_2016,old_gogreen_2020,contini_roles_2020}, our results in Figure \ref{av_mass} may provide more context. When comparing the average stellar masses of our environment samples, we highlighted in \S~\ref{subsec:mass} that the $z_{1}$ and $z_{2}$ high density environments contained, on average, massive ${\rm log(M_{*}/M_\odot)}\geq10.2$ galaxies. Taken at face value, the apparent build-up of stellar mass may indicate that merger interactions are present in $z_{1}$ and $z_{2}$. This hypothesis is supported by \citet{allen_differential_2015}, who suggested star-forming cluster galaxies at $z\sim2.1$ may be experiencing growth via minor mergers, as well as other high redshift studies \citep[e.g. $z\leq3$;][]{conselice_direct_2003,conselice_galaxy_2006,lin_redshift_2008,naab_minor_2009,van_dokkum_growth_2010} who argued that galaxy mergers were the most dominant interaction in the early universe. 

Because we are working with massive systems, however, we highlight that the apparent suppressed star formation activity and large quiescent fractions may be a result of stellar mass quenching and not necessarily environmental-based quenching. \citet{kawinwanichakij_effect_2017}, for example, find that galaxy quenching via environmental processes dominate for lower-mass galaxies (${\rm log(M_{*}/M_\odot)}\leq9.5$) towards lower redshifts ($z<1.5$), while at higher redshifts ($z<2$) environmental and mass quenching processes are indistinguishable. It is plausible to suggest that at our redshifts, environmental and stellar mass quenching processes may have been operating closely, and that the high density environments of $z_{1}$ and $z_{2}$ built up stellar mass via merger interactions while mass quenching is more likely to be responsible for the quiescent and sSFR results of Figures \ref{q_frac} and \ref{av_ssfr}, respectively. \citet{darvish_effects_2016}, who studied the effects of environment and stellar mass on $z<3$ galaxy quenching suggest something similar, finding that while quiescent fractions are dependent on environment at $z\leq1$, they become more dependent on stellar mass out to $z\sim3$. Similarly, \citet{grutzbauch_relationship_2011} argued that very overdense environments tended to have suppressed star formation activity up to $z\sim2$, possibly due to merger activity (though it is worth noting that for the most part, they found that local environment appears to have very little effect on galaxy SFRs at $z>1.5$). The connection between quiescent galaxies, star formation activity, and the role of environmental and stellar mass quenching is clearly challenging to disentangle at higher redshifts. Given previous evidence, we cannot rule out the possibility of hidden stellar mass quenching processes in Figures \ref{q_frac} to \ref{av_ssfr}. In addition, we also need to carry out other tests to investigate whether galaxy mergers are present, which may be achieved by analysing dustiness \citep[e.g.][]{tacconi_submillimeter_2008,casey_dusty_2014}, effective half-light radii \citep[e.g.][]{allen_differential_2015} and possibly gas tracing \citep[e.g.][]{puglisi_titanic_2021}.

It is currently thought that massive galaxy clusters which pervade the local universe \citep[e.g.][]{hernandez-fernandez_disentangling_2012} were assembled via merging of group structures \citep[e.g.][]{press_formation_1974,fakhouri_merger_2010}, according to the $\Lambda$CDM hierarchical models. This means that prior to infall, some group galaxies have already experienced environmental interactions and star formation quenching, more commonly known as \textit{pre-processing} \citep[e.g.][]{fujita_pre-processing_2004,wetzel_galaxy_2013,bianconi_locuss_2018,olave-rojas_galaxy_2018}. Because our study involves high redshift galaxies, we suspect that the high density environments are still in early stages of development and that they will merge with other galaxies and small structures \citep[e.g.][]{spitler_first_2012} before settling into the large clusters that we observe in the low redshift universe. This means that the Butcher-Oemler effect and SFR-density relation may be difficult to pin down here, given that our high density galaxies could undergo positive or negative growth mechanisms (e.g. cold gas accretion and tidal interactions, respectively) if they merge with other structures at later times.

As addressed, the relationship we detect between galaxy evolution and environment appears to be strongest up to $z\sim2.4$. At higher redshifts ($z>3.1$), we suffer from small samples of quiescent sources and sSFR data and express caution with these results. Across our quiescent fraction, stellar mass and sSFR findings (Figures \ref{q_frac}, \ref{av_mass} and \ref{av_ssfr}, respectively), we also encounter large errors and error overlap of some of the samples. Despite this, we established that the high density environments of $z_{6}$ and $z_{8}$ have the largest stellar masses. Our data may suggest that we are detecting some environmental influence across $z_{6}$ and $z_{8}$, and that it may strengthen over time (as observed at $z_{1}$ and $z_{2}$), but a more complete sample is needed to confirm this result.

We recognise that \citep[excepting the H2 and H3 regions of $z_1$;][]{yuan_keckmosfire_2014} the high density environments of this study are photometric candidates and require spectroscopic confirmation for better characterisation, should they be studied in more detail. With deep near-IR imaging from its medium-band filters, the ZFOURGE survey has provided accurate samples of high redshift galaxies that has allowed us to explore the early universe \citep[e.g.][]{spitler_first_2012,cowley_zfourge_2016,cowley_decoupled_2018,kawinwanichakij_satellite_2016,allen_size_2017,papovich_effects_2018}, but remains somewhat limited beyond $z>3.1$. To continue investigating early galaxy environments, we require larger datasets of high redshift sources which may be possible with the new Keck Wide-Field Imager and upcoming observations from the James Webb Space Telescope (JWST). Further investigation of the high density environments probed in this work may be considered as one of the scientific goals of ZFOURGE 2, should a follow-up survey be conducted.

Finally, it is worth highlighting that we observed a very small volume of the universe through ZFOURGE and that cosmic variance may have an impact on our findings. While we detected environmental influence across $z_{1}$, $z_{2}$ and potentially $z_{6}$ and $z_{8}$, we cannot say whether all overdense environments at these redshifts are accelerating galaxy evolution. Even though our results are supported by many previous studies, we acknowledge that any differences between our analyses and others is likely attributed to the phenomenon of cosmic variance, differences between galaxy surveys and environment detection and isolation techniques.

\section{CONCLUSIONS}
\label{sec:summary}
We have studied the effects of environment on galaxy evolution across $2.0\leq z<4.2$ using mass-limited ${\rm log(M_{*}/M_\odot)}\geq9.5$ COSMOS sources from the ZFOURGE survey. We divided our sample into eight redshift slices and searched for low, intermediate and high density environments using the 7NN algorithm. We discovered new overdense candidates across four redshift slices, $2.0\leq z<2.2$ ($z_{1}$), $2.2\leq z<2.4$ ($z_{2}$), $3.1\leq z<3.4$ ($z_{6}$) and $3.8\leq z<4.2$ ($z_{8}$). Our main findings are as follows:

\begin{enumerate}
    \item The $z_{1}$ and $z_{2}$ high density environments exhibit elevated quiescent fractions, contain ${\rm log(M_{*}/M_\odot)}\geq10.2$ massive sources and suppressed star formation activity. Average stellar masses of $z_{1}$ and $z_{2}$ were reported as ${\rm log(M_{*}/M_\odot)}\sim10.220\pm0.096$ and $\sim10.571^{+0.139}_{-0.136}$, respectively. sSFRs decreased from low to high density by $\sim34\%$ and $\sim60\%$ for $z_{1}$ and $z_{2}$, respectively.   
    \\
    \item The results for $z_{1}$ and $z_{2}$ suggest that $z<2.4$ high density environments already consist of evolved, passive galaxies. The significant build-up of stellar mass may also be indicative of merger processes.
    \\
    \item With elevated stellar masses, we may have evidence to suggest that the $z_{6}$ and $z_{8}$ high density environments have begun to influence galaxy evolution, but require more complete samples to confirm this finding.  
\end{enumerate}

While the primary quenching mechanism is difficult to disentangle, our findings suggest that there is a correlation between high redshift galaxy evolution and environment. With our results also implying that the Butcher-Oemler effect and SFR-density relation may not reverse at higher redshifts, it seems that the role of environment in galaxy evolution continues to be a challenge requiring further investigation.

\begin{acknowledgement}
We thank the anonymous referee for constructive comments that improved aspects of the paper. This research includes data collected by the 6.5-metre Magellan Telescopes at the Las Campanas Observatory, Chile. Data analysis was carried out using the Python 3 coding language. Packages of Python that were employed in this paper include matplotlib, a library for creating high quality graphs, Astropy, a community-developed package that contains functions for astronomical research, and NumPy, a library for scientific computing. This research received no specific grant from any funding agency, commercial, or not-for-profit sectors.
\end{acknowledgement}


\bibliography{References.bib}



\end{document}